\documentstyle[11pt,epsf]{article}

\begin{document}
\begin {center}

{\Large On the Stability of Stellar Systems with Double Massive Centre}
\vspace{0.4mm}

K.M. Bekarian
\vspace{0.1mm}

{\it Department of Applied Mathematics, Yerevan State University ,
A. Manugian 1, Yerevan, Armenia}
\vspace{0.2mm}

A.A.Melkonian
\vspace{0.1mm}

{\it Theoretical Department, Yerevan Physics Institute, Alikhanyan
Br. str.2, Yerevan, 375036, Armenia}
\end{center}

{\bf Abstract}
The Ricci curvature criterion is used for the investigation
of the relative instability of several configurations of N-body gravitating
systems. It is shown, that the systems with double massive centers are more
unstable than the homogeneous systems and those with one massive center.
In general this shows the efficiency of the Ricci curvature method
introduced by Gurzadyan and Kocharyan (1987)
for the study of N-body systems via relatively simple calculations, i.e.
for small $N$, and hence small computer resources.

\section{Introduction}

The interest to the problem  of dynamics of stellar systems 
with double massive centre
is due to the discoveries
of double nuclei in the centers of several galaxies
have various separation in projection -- from 2 pc for M31 up
to 800pc for Markarian 273; double massive object was descovered in the centre
of Arp 220 \cite{BH} too. A massive binary system located in the core 
of a galaxy can lead to a number of dynamical effects.

In the present paper we investigate the role of double centers, i.e. 
binary massive bodies situated in the center of N-body systems,
in the dynamical instability of the gravitating systems using  the Ricci
criterion (has been introduced in \cite{GK1}) for estimation of  the
relative instability (chaos) of those
systems. 
This criterion, for example, had enabled to establish in
\cite{GKDAN} that the regular
central field is increasing the instability of the N-body systems.
It is remarkable that this result has been obtained via numerical study of
relatively small number of particles and later has been confirmed
with extensive simulations on powerful computers \cite{GZ}.
The effect of the central regular field is crucial for the
understanding of the relaxation, mixing and evolution of the galactic
cores.
The situation is complicated in a sense that there are
both type of effects - acting to increase and
decrease the chaos in the system.

The flow of numerical methods range from
the Lyapunov numbers and KS-entropy, approximate expansion and
frequency maps up to powerful methods based on direct solutions
\cite{GW}, \cite{A1}, \cite{LL}.
The adequacy and efficiency of a given method in each particular
case is itself an interesting problem.
For example, the efficiency of estimation of Lyapunov numbers is 
limited by a number of reasons, particularly due to
the exponential growth of the errors at large enough N
inevitable at any iterated numerical procedure.

Geometrical methods based on the theorems of the theory of dynamical systems
provide an alternative way of study of N-body systems
via reducing the problem to that of the geometry of the
phase and configurational spaces of the system \cite{Arn}.
In physical problems this method has been initially
used by Krylov \cite{Krylov} and for N-body
gravitational systems -- by Gurzadyan and Savvidy \cite{GS}.
In the latter papers it was shown that spherical N-body systems are
K-systems, i.e. are mixing systems with exponential instability.

The statistical properies plays a key role for the understanding 
of the relaxation and evolution of
many astrophysical objects - from the Solar system to galaxies and
clusters of galaxies.

First, we will briefly describe the Ricci curvature formalism, and
the algorithm of the numerical calculations.

\section{The Ricci curvature criterion}

N-body gravitating system with potential
$$
V=-G\sum_{i<j}^{N} Gm_im_j/r_{ij},
$$
($r_{ij}$ is the distance between the particles with masses $m_i$
and $m_j$) via a variational principle can be transformed to a geodesic
flow in a Riemannian space\cite{Arn}.
The behavior of close geodesics in this space is described by
the Jacobi equation
$$
\nabla_u \nabla_u n+Riem(n,u)u=0,
$$
where $u$ is the velocity of geodesics, $n$ is the separation vector of
close geodesics and $\nabla$ denotes the covariant derivative.
For the normal component
of the deviation one can obtain the following equation

$$
\frac{d^2 \parallel n \parallel ^2}{ds^2}= -2K_{u,n}\parallel n \parallel^2 +2
\parallel \nabla_u n \parallel,
$$

where $$K_{u,n}=  \frac{[Riem(n,u)u]n}{\parallel n \parallel}$$
is the so-called two-dimensional curvature (Riem is the Riemannian curvature).

If $K_{u,n}$ is strongly negative in all
two-directions (u,v) and everywhere in a compact
manifold, the system possesses  maximally strong instability
properties is isomorphic to Bernoulli shift and is an Anosov system \cite{An}.

However this condition is too strong and therefore is not fulfilled
for real physical systems.
It is reasonable therefore to look for a weaker criterion
using some average deviation of
geodesics and a mean curvature in the manifold $M$.
Consider the Ricci curvature

$$
r_u(s)=R_{\mu\nu} \frac{u^{\mu}u^{\nu}}{\parallel u \parallel^2},
$$

where $R_{\mu\nu}$ is the Ricci tensor.

The criterion of relative instability based on the Ricci curvature
reads (Gurzadyan and Kocharian \cite{GK1}):

The geodesics $\gamma_1(s)$ with velocity $u_1$ is more unstable with
respect to the geodesics
$\gamma_2(s)$ with velocity $u_2$ within some interval $[0;S_1]$, if

$$
r= \frac{1}{3N} \rm inf[r_u(s)]
$$

and 

$$
r_1 < r_2; r_1 <0.
$$
The advantage of this criterion is that it is checkable via computer
simulations of many  dimensional systems, including of N-body systems.
Note also, that as distinct from the Lyapunov numbers, this criterion
describes the local (in time) properties of the system. 

\section{Numerical simulations}

\vglue 0.3cm
\begin {figure}[b]
\centering\leavevmode
\epsfxsize=\hsize
\epsfbox {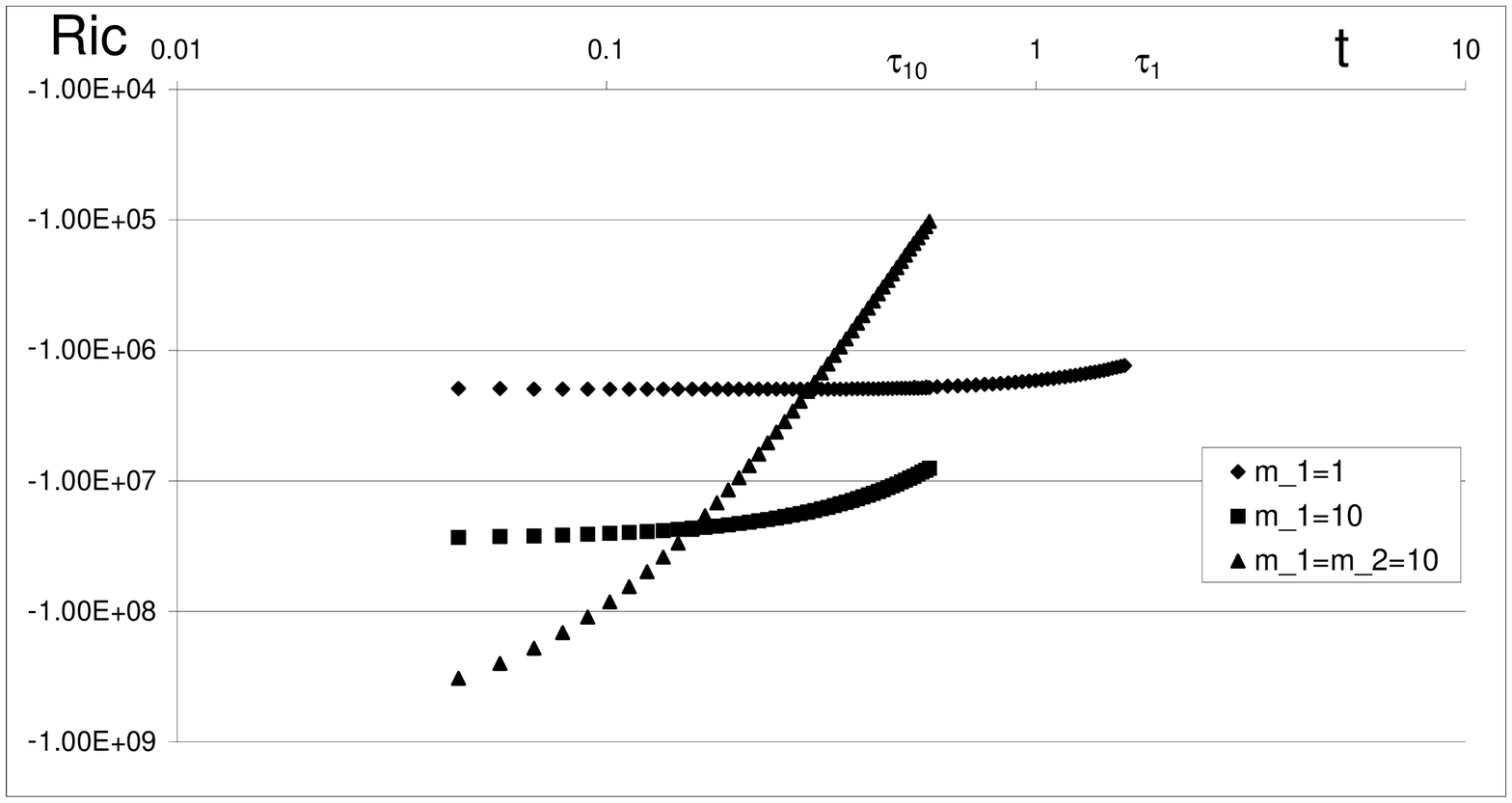}
\vglue -0.3cm
\caption{Comparative instability of three types of 
systems: with equal masses of particles, with a single central mass
$m_1=10$ and with two massive centres $m_1=m_2=10$; here and in the
following figures
$m_1, m_2$ are the masses of the central particle/particles, while the
other $N-1$ particles have $m=1$.}
\end{figure}

\vglue 0.3cm
\begin {figure}[b]
\centering\leavevmode
\epsfxsize=\hsize
\epsfbox {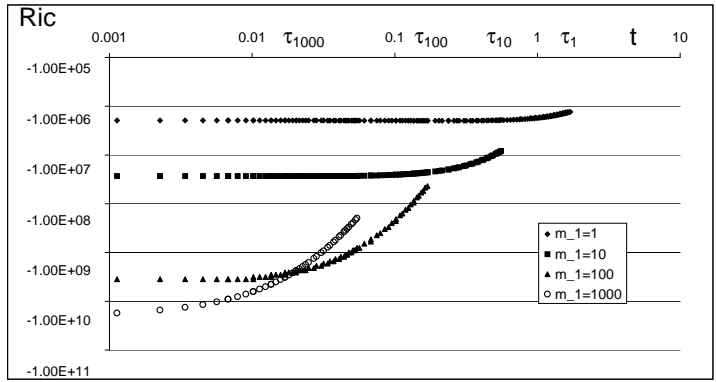}
\vglue -0.3cm
\caption{Comparison of configurations with various parameters of
masses.}
\end{figure}

\vglue 0.3cm
\begin {figure}[b]
\centering\leavevmode
\epsfxsize=\hsize
\epsfbox {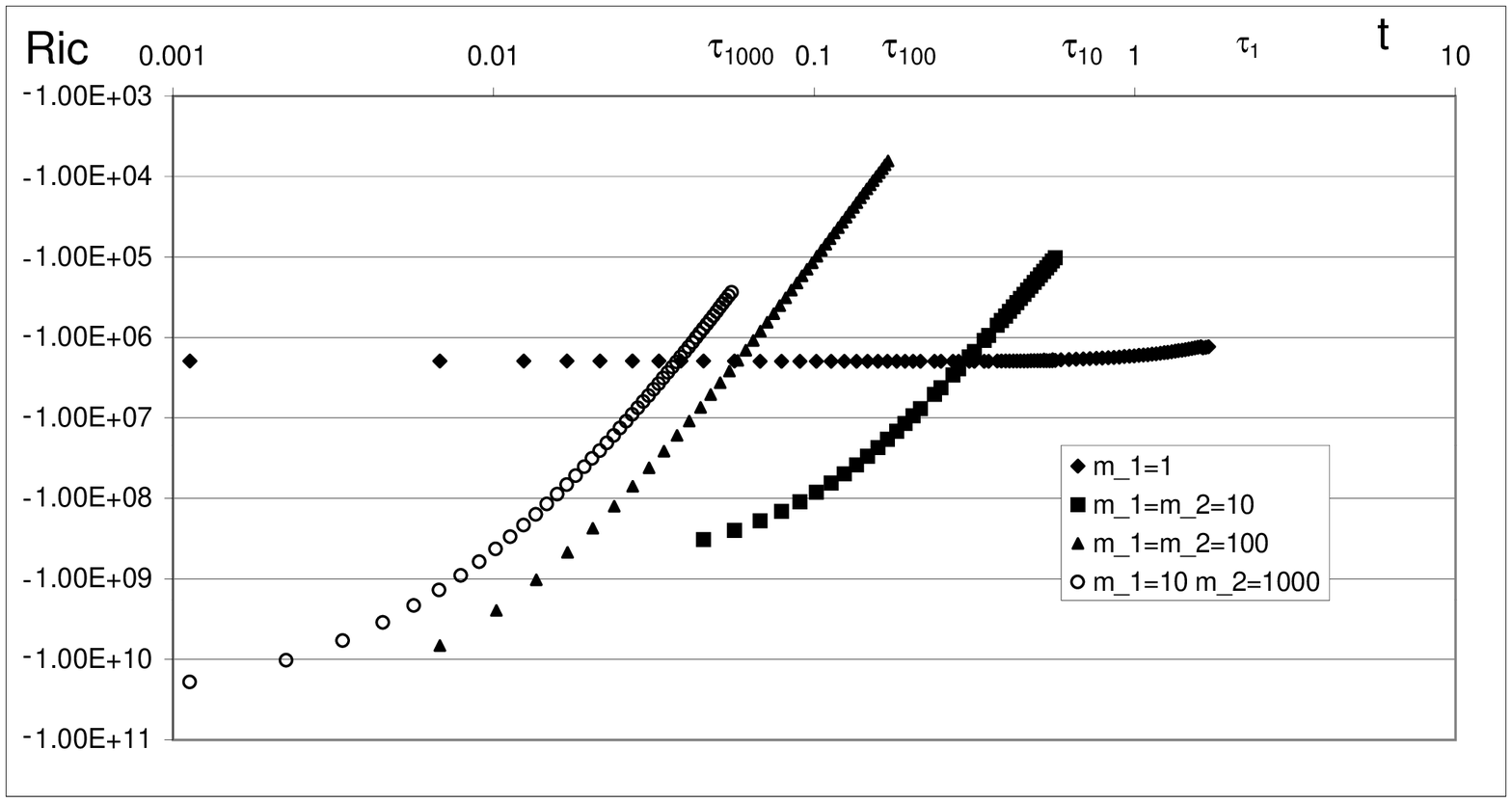}
\vglue -0.3cm
\caption {Comparative instability of systems with higher values of 
central masses.}
\end{figure}

At our computer experiments we estimated the Ricci curvature for
several evolving configurations, i.e. traced the variation
of the Ricci curvature in time.
The formula for the Ricci curvature for N-body systems
can be derived from the formulae given in previous section and
has the following form \cite{GK1}, \cite{GK2}, \cite{GP}.

$$
r_u(s)=- \frac{(3N-2)}{2} \frac{W_{ik}u^iu^k}{W}+ \frac{3}{4}(3N-2)
\frac{(W_iu^i)^2}{W^2}- \frac{(3N-4)}{4} \frac{|\nabla W|^2}{W^3},
$$
where
$$
W=E-V;\; W_i= \frac{\partial W}{\partial q_i};
$$
$$
W_{ik}= \frac{\partial^2 W}{\partial q_i \partial q_k};\;
\Delta W= \sum_{i}W_{ii};
$$
$$
|\nabla W|^2= \sum_{i}(\frac{\partial W}{\partial q_i})^2.
$$

We built the systems with $N=22$ using the scheme
described in \cite{GK1}, i.e.
we considered systems, so that the particles were
located at the apexes of concentric cubes of unit sides.
The velocities of particles were chosen in way to have
no rotational momentum for the system.

We estimated the variation of the Ricci curvature by time 
for the following configurations (Figures 1-3):

1. Homogeneous, i.e. all particles have the same mass $m$;

2. With one central massive particle $m_1$, while $N-1$ particles
have the same mass $m<<M$;

3. With two massive particles of masses $m_1$ and $m_2$
situated in the central part of the system, while $N-2$
particles have the same mass $m<<m_1,m_2$;.

Our calculations showed that the systems with massive
center are more
unstable than the homogeneous ones, thus confirming the conclusions in
\cite{GK1}, \cite{GKDAN}, \cite{GZ}, \cite{Z}.
Note, the growth of the instability with the increase of the central mass $M$.

Figures 1-3 illustrate the comparative instability of the three
different types of systems
mentioned above. The most unstable among the
initial configurations is the system with double-central masses. 
Note, that as the systems evolves the Ricci curvature
tends to zero for all three systems, however the rate of
tending is most rapid for the system with double centers.
Physically it is clear that the third type of systems has to
dissolve quicker for small number of particles. Just this tendency
has been also noticed at the numerical experiments, namely the tending to
zero of the Ricci curvature becomes slower with the increase of $N$.

In other words, the double massive central objects make the system
more unstable initially, however then the system evolves quicker to its final
dissolved state with regular orbits.

This tendencies have been confirmed in numerous
experiments with configurations with various initial conditions.

\section{Conclusions}

Thus we used the Ricci curvature criterion to study the
relative instability of three types of N-body systems:
homogeneous systems, those with one central mass and two
central massive bodies. The following main conclusions had been
drawn via the numerical experiments:
 
1. The presence of the second massive central object makes the
system more unstable as compared with that those of a
single massive center and 
the homogeneous ones.

2. The system with double massive objects is evolving more quicker
towards dissolution, i.e. to a more global regular situation.

3. The greater is the ratio of the mass of the massive objects to the
mass of the rest particles, the more is the difference
in the initial instability and the rate of evolution.

The main conclusion is however the efficiency of the
Ricci curvature criterion for the study of such complex
many dimensional systems -- N-body systems -- by means of
simple numerical experiments.

We  thank V.G.Gurzadyan and S.J.Aarseth for valuable discussions. 
This work is supported in part by INTAS grant.

\end{document}